\documentclass[jcp, floatfix, nobibnotes, reprint, superscriptaddress]{revtex4-1}
\usepackage{docs}%
\usepackage{siunitx}
\usepackage{amsmath}
\usepackage{booktabs}
\DeclareSIUnit[number-unit-product = {\,}]
\cal{cal}
\DeclareSIUnit\kcal{\kilo\cal}
\DeclareSIUnit[number-unit-product = {\,}]
\Btu{Btu}
\DeclareSIUnit[number-unit-product = {\,}]
\Fahr{\degree F}
\DeclareSIUnit[number-unit-product = {\,}]
\usepackage{bm}%
\usepackage[colorlinks=false,linkcolor=black]{hyperref}%
\usepackage{graphicx}
\usepackage[utf8]{inputenc}
\usepackage{tabularx} 

\usepackage{dcolumn}
\usepackage{epstopdf}
\usepackage{afterpage}
\usepackage{setspace}
\usepackage{multirow}
\usepackage{dsfont}
\usepackage{sidecap}
\usepackage{epstopdf}
\usepackage{braket}
\usepackage{csquotes}
\usepackage{cleveref}

\usepackage{siunitx} 
\usepackage{color, colortbl}

\definecolor{Gray}{gray}{0.9}

\AppendGraphicsExtensions{.gif}

\DeclareMathAlphabet\mathbfcal{OMS}{cmsy}{b}{n}

\setcitestyle{super}

\begin{document}
\setstretch{1.0}

 \title[]{Adaptive atomic basis sets}

\author{Danish Khan}
\thanks{Both authors contributed equally}
\affiliation{Chemical Physics Theory Group, Department of Chemistry, University of Toronto,
St. George Campus, Toronto, ON, Canada}
\affiliation{Vector Institute for Artificial Intelligence, Toronto, ON, M5S 1M1, Canada}

\author{Maximillian L. Ach}
\thanks{Both authors contributed equally}
\affiliation{Department of Physics, University of Toronto, St. George Campus, Toronto, ON, Canada}
\affiliation{Department of Physics, Ludwig-Maximilians-Universität München (LMU), Munich, Germany}
\author{O. Anatole von Lilienfeld}
\email{anatole.vonlilienfeld@utoronto.ca}
\affiliation{Chemical Physics Theory Group, Department of Chemistry, University of Toronto,
St. George Campus, Toronto, ON, Canada}
\affiliation{Vector Institute for Artificial Intelligence, Toronto, ON, M5S 1M1, Canada}
\affiliation{Acceleration Consortium, University of Toronto, Toronto, ON, Canada}
\affiliation{Department of Physics, University of Toronto, St. George Campus, Toronto, ON, Canada}
\affiliation{Department of Materials Science and Engineering, University of Toronto,
St. George Campus, Toronto, ON, Canada}
\affiliation{Machine Learning Group, Technische Universit\"at Berlin and Institute for the Foundations of Learning and Data, 10587 Berlin, Germany}
\affiliation{Berlin Institute for the Foundations of Learning and Data, 10587 Berlin, Germany}
\begin{abstract}
\vspace{1mm}
\section*{Abstract}
\vspace{-2mm}
Atomic basis sets are widely employed within quantum mechanics based simulations of matter. 
We introduce a machine learning model that adapts the basis set to the local chemical environment of each atom, 
prior to the start of self consistent field (SCF) calculations.
In particular, as a proof of principle and because of their historic popularity, we
have studied  the Gaussian type orbitals from the Pople basis set, i.e. the STO-3G, 3-21G, 6-31G and 6-31G*. 
We adapt the basis by scaling the variance of the radial Gaussian functions leading to contraction or expansion of the atomic orbitals.
A data set of optimal scaling factors for C, H, O, N and F were obtained by variational minimization of the Hartree-Fock (HF) energy of the smallest 2500 organic molecules from the QM9 database.
Kernel ridge regression based machine learning (ML) prediction errors of the change in scaling decay rapidly with training set size, typically reaching less than 1\% for training set size 2000. 
Overall, we find systematically lower variance, and consequently the larger training efficiencies, when going  from hydrogen to carbon to nitrogen to oxygen. 
Using the scaled basis functions obtained from the ML model, we conducted HF calculations for the subsequent 30'000 molecules in QM9. 
In comparison to the corresponding default Pople basis set results we observed improved energetics in up to $99\%$ of all cases. 
With respect to the larger basis set 6-311G(2df,2pd), atomization energy errors are lowered on average by $\sim$31, 107, 11, and 11 kcal/mol for STO-3G, 3-21G, 6-31G and 6-31G*, respectively --- with negligible computational overhead.
We illustrate the  high transferability of  adaptive basis sets for larger out-of-domain molecules  relevant to  addiction, diabetes, pain, aging. 
Generally, deviation from cc-pvQZ based HF results for counter-poise corrected atomization energies, HOMO-LUMO gaps, and dipole moments also decreases --- in line with the variational principle.
\end{abstract}
\maketitle

\section{Introduction}
The choice of atomic basis set used for expanding molecular orbitals is one of the central components of any quantum chemical calculation due to its large influence on both computational cost and accuracy\cite{jensen_review_2013}.
Since their popularization in the early 1970s, Gaussian Type Orbitals (GTOs) have been by far the most widely used type\cite{sto-ng, 4-31g}.
The primary reason for this is the more efficient evaluation of four-center integrals due to the Gaussian product rule.
Since the quality of the wavefunction and the cost of a quantum mechanical calculation directly depend on the size and quality of the basis set, several strategies and basis set families have been developed to improve this trade-off.
A detailed documentation can be found in numerous, comprehensive reviews\cite{jensen_review_2013,Hill2012, davidson1986basis}.
Since optimizing the basis set individually for each molecule is impractical, the free parameters, namely the contraction coefficients and the radial exponents, are in the majority of cases obtained by fitting to atomic calculations and kept constant for each element\cite{321g,4-31g,631g,dunning1989gaussian,ahlrichs_opt}.
Despite the large success of this approach, it is known that optimal values can differ significantly for molecular systems\cite{4-31g,szabo_ostlund}.
While using such immutable basis sets generally does not pose a problem when large (and balanced) basis sets of triple zeta quality or better are employed, for more incomplete (and frequently used due to tractability) basis sets there is still room for improvement.
\\
In analogy to the recently introduced machine learning (ML) assisted adaptive hybrid density functional approximations\cite{khan2024adaptive}, we introduce ML models that provide improved basis sets (STO-3G\cite{sto-ng}, 3-21G\cite{321g}, 6-31G\cite{631g}, 6-31G*\cite{631gs}
) by adapting them, on-the-fly, to the system of interest.
This is done by scaling the radial exponents of contracted GTOs (CGTOs) by a constant factor depending on the local chemical environment of each atom.
A training dataset of optimal scaling factors is obtained via variational minimization of Hartree-Fock (HF) energies of 2,000 smallest molecules from the QM9\cite{qm9} dataset.
The data is subsequently used to train a ML model of the optimal scaling factors for each atomic basis set.
For out of sample molecules, this leads to improved HF total and atomization energies in up to $\sim$99.9$\%$ of cases at no additional computational cost.
Since the optimal scaling factors are obtained via application of the variational principle, the improved basis sets lead to improved approximations of the exact ground state HF wavefunctions.
As a result of this, simlutaneous improvements are also obtained for other physical properties like HOMO-LUMO gaps and dipole moments indicating a foundational ML model for quantum chemistry. 
Due to the strong locality of the scaling factors, the model is shown to be easily transferable to larger systems of biologically relevant molecules suggesting modest training data needs to be generally applicable.
\section{Theory and methods}
\subsection{Basis function scaling}
Gaussian type basis functions are composed of linear combinations of primitive Gaussian type orbitals (PGTOs) of the form
\begin{align}
    g_{ik}^\text{PGTO}(\alpha_{ik}) &= N_{ik, n}Y_{lm}(\theta,\phi)|\mathbf{r}-\mathbf{R}_{A}|^{n-1} \mathrm{e}^{(-\alpha_{ik}|\mathbf{r}-\mathbf{R}_{A}|^{2})},
    \\
    N_{ik, n} &= \sqrt{\frac{2(2\alpha_{ik})^{n+\frac{1}{2}}}{\Gamma(n+\frac{1}{2})}} \nonumber
\end{align}
which form contracted orbitals (CGTOs) 
\begin{align}
    \phi_{k}^\text{CGTO} = \sum_{i} c_{ik} ~g_{ik}(\alpha_{ik})
    \label{cgto}
\end{align}
where $Y_{lm}$ are spherical harmonics, $\mathbf{R}_{A}$ denotes the nuclear coordinate on which the PGTO/CGTO is centered, $c_{ik}$ are fixed (pre-optimized) linear combination coefficients (called contraction coefficients), $\Gamma$ denotes the gamma function and the sum runs over the length of contraction.
To make these basis functions adaptive to the local atomic environment, the radial exponents of all PGTOs within a CGTO are multiplied by a single constant factor $\zeta_{k}$, i.e.
\begin{align}
    g_{ik, \mathrm{scaled}}^\mathrm{PGTO} (\zeta_{k} \alpha_{ik}) = &N_{ik,nlm}Y_{lm}(\theta,\phi)R_{l}(\mathbf{r},\mathbf{R}_{A})\\&\cdot\mathrm{e}^{(-\zeta_{k}\alpha_{ik}|\mathbf{r}-\mathbf{R}_{A}|^{2})}
    \notag
\label{eq:pgto_scaled}
\end{align}
and
\begin{align}
    \phi_{k, \mathrm{scaled}}^\mathrm{CGTO} = \zeta_{k}^{\frac{3}{2}} \sum_{i} c_{ik} ~g_{ik}(\zeta_{k} \alpha_{ik})
\end{align}
where the prefactor is for normalization. 
$\zeta_{k}$ are unique for each CGTO and assumed to depend on the atom's local chemical environment.
This is similar to the non-adaptive scaling suggested for the 4-31G basis set\cite{4-31g}.
The local atomic environment is defined as the sphere $S_{A}(\mathbf{R}_{A}, r_{\mathrm{cut}})$ of radius $r_{\mathrm{cut}}$ centered at $\mathbf{R}_{A}$.
The $\zeta_{k}$ are then assumed to be functions of the atomic species present within the sphere $S_{A}$, $\zeta_{k} = f(\{Z_{i}, \mathbf{R}_{i}\}_{i:\mathbf{R}_{i} \epsilon S_{A}})$.
The function $f$ in our work is estimated using ML.
In our work we modify valence CGTOs and polarization functions and use the same scaling factors for s and p orbitals, since the exponents between these are shared in the default basis sets as well.
We note that including more scaling factors could however further improve the method.
\\
The optimal set of scaling factors for a molecule, $\boldsymbol{\zeta}_{\text{opt}}$, is obtained by minimizing the HF energy as a function of these $\zeta$
\begin{align}
    \boldsymbol{\zeta}_{\text{opt}} = \arg\min_{\boldsymbol{\zeta}} E_{\text{HF}}(\boldsymbol{\zeta})
\end{align}

Gradients of the HF energy with respect to $\zeta_{k}$ can be calculated using the general HF response expression
\begin{align}
    \frac{\partial E_\text{HF}}{\partial \phi_{k}} &= \sum_{\mu\nu}\biggl(P_{\mu\nu}\frac{\partial h_{\mu\nu}}{\partial \phi_{k}} - W_{\mu\nu}\frac{\partial S_{\mu\nu}}{\partial \phi_{k}}\biggr) \nonumber 
    \\
    &+ \frac{1}{2}\sum_{\mu\nu\lambda\sigma}P_{\mu\nu}P_{\lambda\sigma}\frac{\partial(\mu\nu||\lambda\sigma)}{\partial \phi_{k}}
    \nonumber
    \\
    &= 2\sum_{\nu}\biggl(P_{k\nu}\frac{\partial h_{k\nu}}{\partial \phi_{k}} - W_{k\nu}\frac{\partial S_{k\nu}}{\partial \phi_{k}}\biggr)\\ &+ \sum_{\nu\lambda\sigma}P_{k\nu}P_{\lambda\sigma}\frac{\partial(k\nu||\lambda\sigma)}{\partial \phi_{k}}
\end{align}
and
\begin{align}
    \frac{\partial \phi_{k}}{\partial \zeta_{k}} = \frac{3}{2} \frac{\phi_{k}}{\zeta_{k}} - \zeta_{k}^\frac{3}{2} |\mathbf{r}-\mathbf{R}_{A}|^{2} \sum_{i} c_{ik} \alpha_{ik}~g_{ik}(\zeta_{k} \alpha_{ik})
\end{align}
where $P_{\mu \nu}$, $S_{\mu \nu}$, $W_{\mu \nu}$, $h_{\mu \nu}$ denote elements of the density, overlap, energy weigthed density and core Hamiltonian matrices respectively\cite{ahlrichs_opt}.
It should be noted that an additional term due to $\zeta_{k}$ will appear when gradients such as atomic forces are evaluated.
To ensure smoothness of the forces, the following form can be used for changing geometries:
\begin{align}
    \zeta_{k} &= 1+f(\{Z_{i}, \mathbf{R}_{i}\}_{i:\mathbf{R}_{i} \epsilon S_{A}}) \sum_{j=1}^{n} f_{\mathrm{cut}}(r_{jA})
    \\
    f_{\mathrm{cut}}(r_{jA}) &= 
    \begin{cases}
        \frac{1}{2}\left( \cos(\frac{\pi r_{jA}}{r_{\mathrm{cut}}})+1\right), ~r_{jA}<r_{\mathrm{cut}}\\
        0, ~r_{jA}>r_{\mathrm{cut}}
    \end{cases}  
\end{align} 
where $f_{\mathrm{cut}}$ is a soft cut-off function that smoothly decays to 0 at $r_{\mathrm{cut}}$\cite{acsf} and $n$ denotes the number of atoms in the system.
Nuclear gradients can be evaluated as $\nabla_{\mathbf{R}_{A}}\zeta_{k}\partial_{\zeta_{k}}\phi_{k}\partial_{\phi_{k}}E_{HF}$ using eq. 7 and 8.  along with
\begin{align}
  \nabla_{\mathbf{R}_{A}}\zeta_{k} = \nabla_{\mathbf{R}_{A}}f(\{Z_{i}, \mathbf{R}_{i}\}_{i:\mathbf{R}_{i} \epsilon S_{A}}) \sum_{j=1}^{n} f_{\mathrm{cut}}(r_{jA})\nonumber \\
  - \frac{1}{2}f(\{Z_{i}, \mathbf{R}_{i}\}_{i:\mathbf{R}_{i} \epsilon S_{A}})\sum_{j:\mathbf{R}_{j} \epsilon S_{A}} \frac{\pi\sin(\frac{\pi r_{jA}}{r_{\mathrm{cut}}})}{r_{\mathrm{cut}}r_{jA}}(\mathbf{R}_{A} - \mathbf{R}_{j})
\end{align}
where the term $\nabla_{\mathbf{R}_{A}}f(\{Z_{i}, \mathbf{R}_{i}\}_{i:\mathbf{R}_{i} \epsilon S_{A}})$ needs to be evaluated by propagating the gradient through the ML model\cite{op_response_anders}.
\\
In the current work we have employed numerical gradients (central finite differences with step size $10^{-8}$) for simplicity. 
The L-BFGS\cite{lbfgs} optimizer, as implemented in Scipy\cite{virtanen2020scipy}, was used to obtain optimal $\zeta_{k}$ of the training set molecules by minimizing HF total energies.
HF calculations were performed using version 2.3.0 of the PySCF package\cite{pyscf1, pyscf2}.
All basis sets were obtained from basis set exchange\cite{basis_set_exchange_bse}.
\begin{figure*}[htbp]
    \centering
    \includegraphics[width=\linewidth]{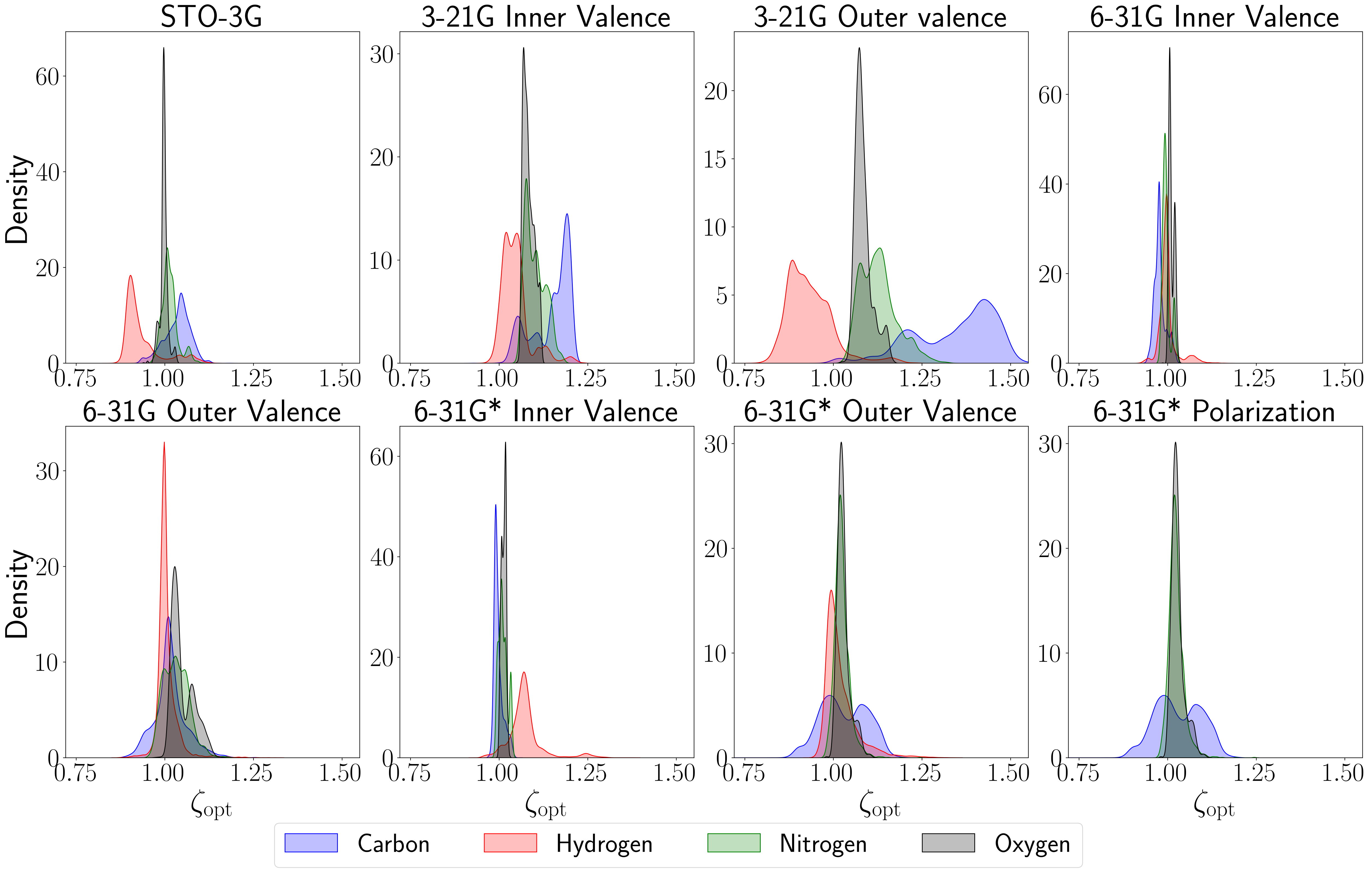}
    \caption{Optimal scaling factor, $\zeta_{\mathrm{opt}}$, distributions for Hydrogen, Carbon, Nitrogen and Oxygen valence shell orbitals from the first 2000 smallest QM9\cite{qm9} molecules found via minimization of total HF energies employing numerical gradients.
    Data used for training, see learning curves in Fig 2.}
\label{fig:Scaling_Dist1}
\end{figure*}
Using this methodology, we calculated optimal scaling factors of the first 2000 molecules from QM9\cite{qm9}, a dataset of 134k small organic molecules with up to nine heavy atoms out of C,O,N,F. 
For STO-3G, this was additionally done for 30k more molecules.
The distribution of $\zeta_{\mathrm{opt}}$ can be seen in Fig. \ref{fig:Scaling_Dist1} for each chemical element and contracted GTO.

\subsection{Machine Learning}
The ML model used throughout this work is Kernel Ridge Regression\cite{vapnik1999nature, rasmussen2006gaussian, GPR_deringer} (KRR) due to its high robustness and simplicity, which has been covered extensively\cite{felix_google,bing_DMC,m3l,siwoo,khan2024adaptive}.
Using this method, the function $f$ is approximated as a weighted sum of similarity measures to all atoms (of the same type) in the training set
\begin{align}
    f(\{Z_{i}, \mathbf{R}_{i}\}_{i:\mathbf{R}_{i} \epsilon S_{A}}) = \sum_{J}^{N_{\mathrm{train}}} \alpha_{J}k(\mathbf{x}(\{Z_{i}, \mathbf{R}_{i}\}_{i:\mathbf{R}_{i} \epsilon S_{A}}), \mathbf{x}_{J})
\end{align}
where $\alpha_{J}$ are the regression weights, $\mathbf{x}$ is the atomic representation feature vector which depends only on the set of nuclear charges and coordinates within the atom's local environment $S_{A}$ ($\{Z_{i}, \mathbf{R}_{i}\}_{i:\mathbf{R}_{i} \epsilon S_{A}}$), and $k(.,.)$ denotes a kernel function acting as a similarity measure.
The kernel function used in our work is the atomic Gaussian kernel
\begin{align}
    k(\mathbf{x}_I, \mathbf{x}_J) = \exp{\left( -\frac{\vert \vert \mathbf{x}_I - \mathbf{x}_J \vert \vert^{2}_{2}}{2 \sigma^2} \right)}
    \label{eq:kernel_gaussian}
\end{align}
The gradient $\nabla_{\mathbf{R}_{A}}f(\{Z_{i}, \mathbf{R}_{i}\}_{i:\mathbf{R}_{i} \epsilon S_{A}})$ in eq. 11 is then straightforward to evaluate as 
\begin{align}
     \nabla_{\mathbf{R}_{A}}f = -\nabla_{\mathbf{R}_{A}} \mathbf{x}_{I} \sum_{J}^{N_{\mathrm{train}}} \alpha_{J} \frac{\vert \vert \mathbf{x}_I - \mathbf{x}_J \vert \vert_{2}}{\sigma^2} k(\mathbf{x}_{I}, \mathbf{x}_{J})
\end{align}
The regression weights $\mathbf{\alpha}$ are obtained from the set of training labels $\mathbf{y}^{\mathrm{train}}$ via the following equation

\begin{equation}
    \mathbf{\alpha} = (\mathbf{K} + \lambda \cdot \mathbf{I})^{-1} \mathbf{y}^{\mathrm{train}}
    \label{eq:solution_alpha}
\end{equation}

where $\mathbf{K}$ is the kernel matrix of the training set and $\lambda$ is a regularization parameter.
\\
In our work the training and predictions are performed separately for each unique chemical element.
This leads to improved predictions since the basis set scalings have different variances for each chemical element (cf. Fig. \ref{fig:Scaling_Dist1}).
Moreover, this also leads to a reduction of the size of the kernel matrix $\mathbf{K}$ from $(\sum_{i} n_{i})^2$ to $(\max n_{i})^2$, where $n_{i}$ denotes number of atoms belonging to chemical element $i$ in the training set.
This results in more efficient training and predictions since the computational cost and memory requirement of the method scale with the size of $\mathbf{K}$.

In this work we have relied upon an extended version of the Many-Body Distribution Functionals (MBDF) representation for generating atomic feature vector mappings $\mathbf{x}$\cite{mbdf}.
The reason for this choice is its compact size, along with high predictive power, which makes the impact on the quantum chemistry calculation cost negligible.
Briefly, MBDF generates the feature vector components as suitable functionals of two- and three-body distributions:
\begin{align}
    F_{\textrm{2}}^{nm}[i]&=\int_{0}^{\infty}dr~g_{n}(r) ~\partial_{r}^{m}\rho_{i}(r),
    \\
    \rho_{i}(r) &= \sum_{j:\mathbf{R}_{j} \epsilon S_{i}}^{M} ~Z_{j}\mathcal{N}{\left(R_{ij},\sigma_{r}\right)}
    \nonumber
\end{align}
\begin{align}
    F_{\textrm{3}}^{nm}[i] &=\int_{0}^{\pi}d\theta~g_{n}(\theta) ~\partial_{\theta}^{m}\rho_{i}(\theta),
    \\
    \rho_{i}(\theta) &= \sum_{j:\mathbf{R}_{j} \epsilon S_{i},k:\mathbf{R}_{k} \epsilon S_{i}}^{M} ~\frac{(Z_{j}Z_{k})^\frac{1}{2}\mathcal{N}\left(\theta_{ijk},\sigma_\theta\right)}{(R_{ij}R_{jk}R_{ik})^2}
    \nonumber
\end{align}
where $\mathcal{N}$ denotes a normalized Gaussian, $Z_{j}$ corresponds to the nuclear charge of atom $j$, $R_{ij}$ and $\theta_{ijk}$ denote interatomic distances and angles respectively and $g_{n}$ are suitable weighting functions similar to 2, 3-body interaction potentials. 
Gradients $\nabla_{\mathbf{R}_{A}} \mathbf{x}_{I}$ of the generated feature vectors (eq. 14) are implemented in the MBDF code\cite{mbdf} (cf. Data and code availability).

\section{Results and Discussion}
\label{sec:Results}
\subsection{Numerical results}
As outlined above, optimal scaling factors were calculated for the first 2500 molecules from the QM9\cite{qm9} dataset, and then randomly drew 500 for testing and up to 2000 of the remaining for training.
The data set comprises molecules with a 14669 hydrogen, 8860 carbon, 2356 oxygen, 1713 nitrogen and 22 fluorine atoms.
Scaling factors for the latter varied only negligibly and were hence neglected for the rest of the study. 
During the optimization, we generally found $\zeta$ to be well behaved and convex, making it suitable for machine learning.
Further, we note that a larger and better sampled training set is likely to make the model even more transferable through the use of atomic fragments based on amons\cite{amons_slatm}.
\\
\begin{figure*}[!htbp]
          \centering           
          \includegraphics[width=\linewidth]{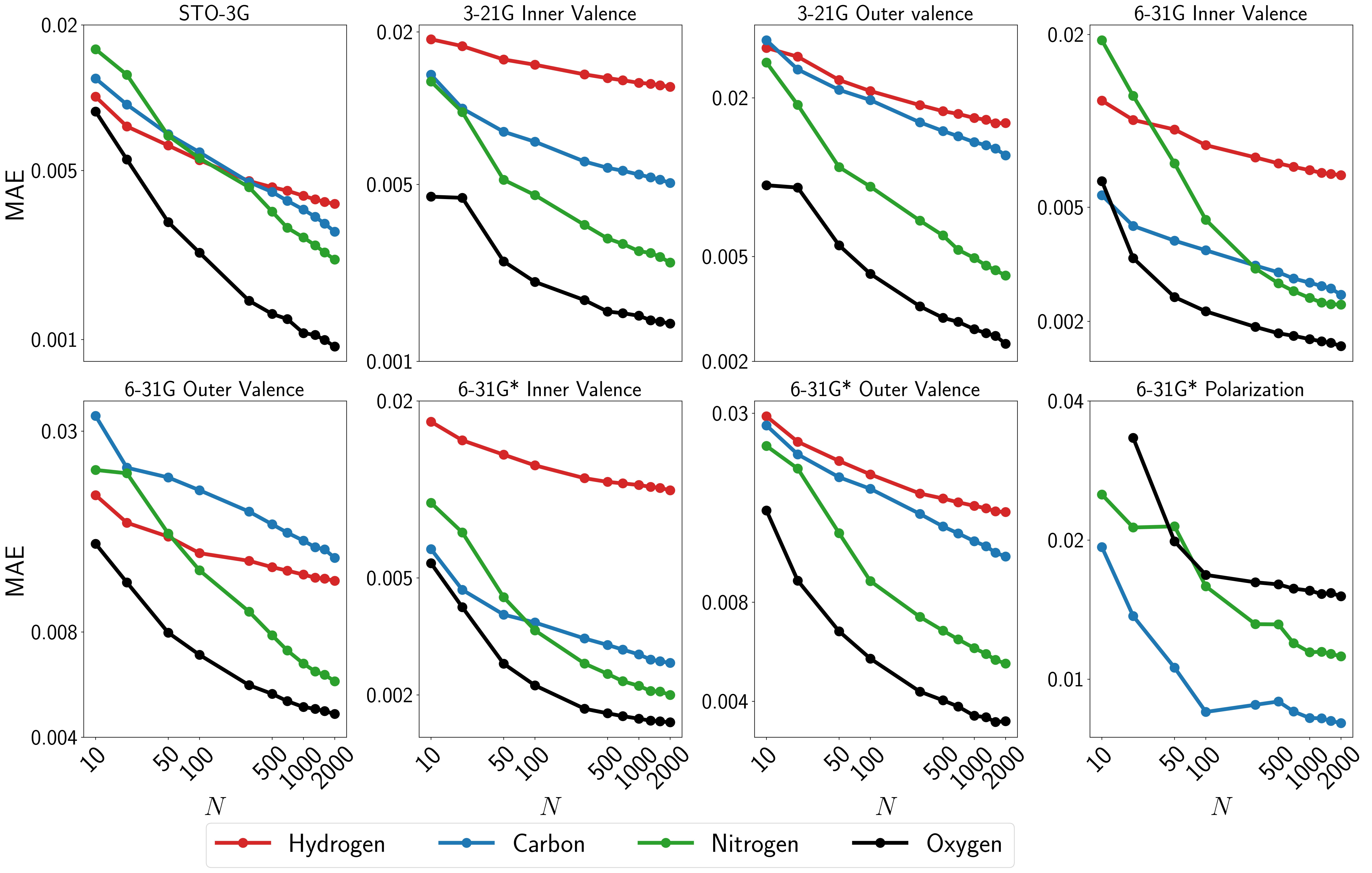}
          \caption{Mean absolute errors (MAE) of predicted scaling factors $\zeta$ for STO-3G (valence orbitals), 3-21G, 6-31G (both inner and outer valence) and 6-31G* (inner and outer valence and polarization functions) as a function of training set size (number of molecules) using the KRR-MBDF model on a validation set of 500 out-of-sample QM9 molecules.
          Note that the average number of atoms per molecule is different for each chemical element (7.33 Hydrogen, 4.43 Carbon and $\sim$1 Nitogen/Oxygen atoms per molecule).
          Training sets were obtained from the $\zeta_{\mathrm{opt}}$ values calculated for the first 2000 smallest QM9 molecules.
          Refer to Fig.~\ref{fig:Scaling_Dist1} for training data distributions.
          }
     \label{fig:lc}
 \end{figure*}
The scaling factors were found to be machine-learnable quite efficiently using our very simple KRR-MBDF model as shown in the learning curves in Fig. \ref{fig:lc} for a set of 500 out-of-sample molecules from the QM9 dataset.
Lower offsets for Nitrogen and Oxygen were generally observed throughout likely due to the small variance of $\zeta_{\mathrm{opt}}$ values (Fig. 1) for these elements.
Larger offset and more erratic behaviour for the 6-31G* polarization function learning is likely because there are 3 scaling factors in this basis set per atom which are coupled to each other whereas the ML models are separate for each.
Distributions in Fig. 1 indicate that variance of the optimal scaling factor is inversely proportional to the electronegativity of the element which is in line with chemical intuition.
\begin{figure}[!htbp]
          \centering           
          \includegraphics[width=\columnwidth]{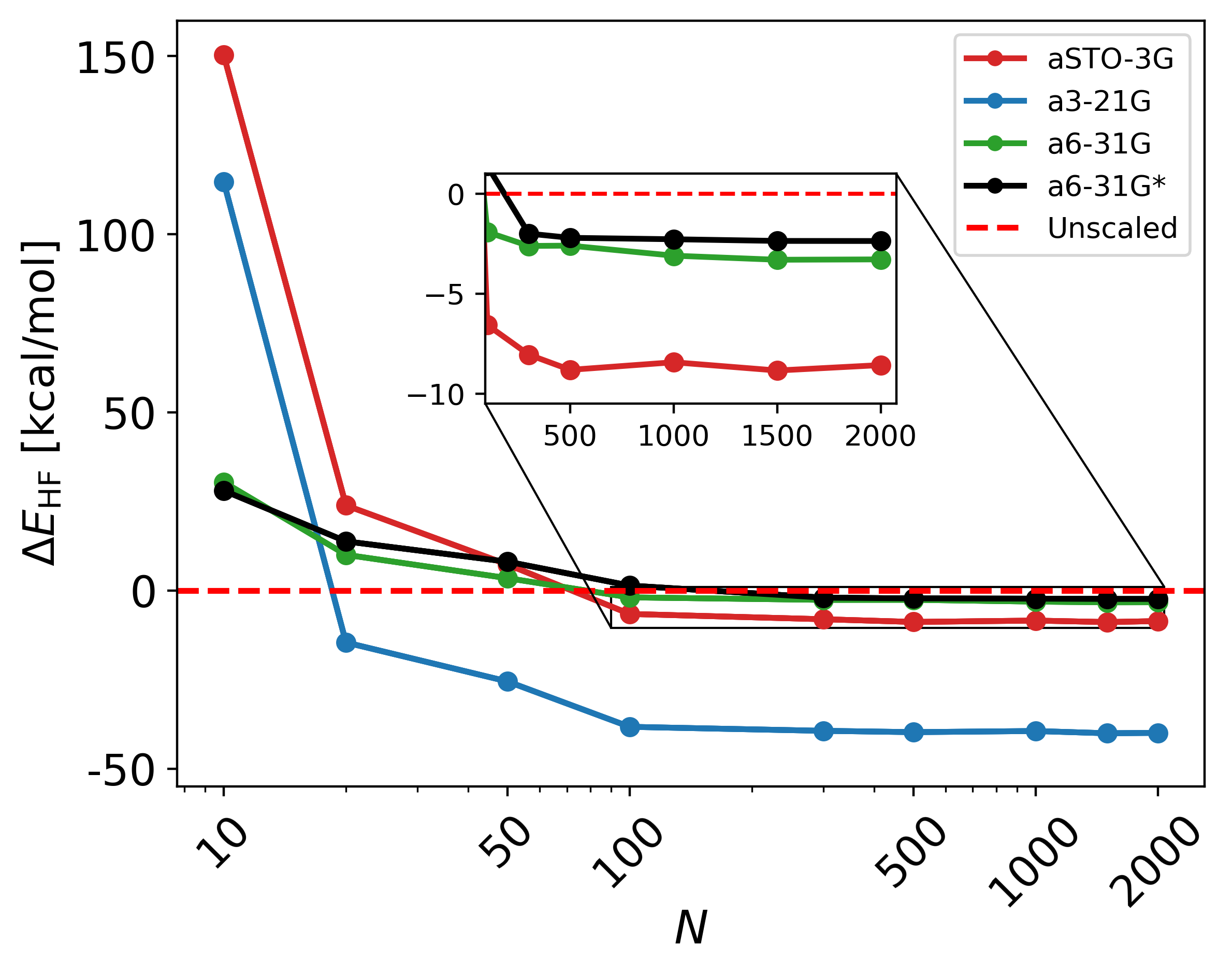}
          \caption{Change in Hartree Fock total energy for a validation set of 500 QM9\cite{qm9} molecules compared to the default unscaled form as a function of training set size.}
     \label{fig:perf_curve}
 \end{figure} 
This indicates that the method should be quite transferable with modest training data needs for such elements.
For Carbon and Hydrogen as well we found the prediction errors to be sufficiently small so as to lead to considerable improvements for out-of-sample molecules.
This was verified by using the ML model to predict basis sets adapted to 30,000 larger molecules drawn at random from the QM9 dataset.
Fig.~\ref{fig:perf_curve} shows changes in total energies obtained from HF calculations when using the adaptive basis sets as a function of training set size.
Remarkably, it can be seen that the adaptive basis sets already start improving upon the default Pople basis sets with $\sim$100 training molecules and only $\sim$500 molecules are sufficient to recover the vast majority of the possible accuracy increase.

\begin{table*}[!htbp] 
    \centering
    \scalebox{0.8}{
    \hspace{-3mm}
    \renewcommand{\arraystretch}{1.5}
    \begin{tabular}{|l|c|c|c|c|c|c|}
        \hline
        & \multicolumn{2}{c|}{\textbf{Total Energies}} & \multicolumn{2}{c|}{\textbf{Atomization Energies}} & \textbf{Average CPU time}\\
        \hline
        \textbf{Basis Set} & \multicolumn{1}{p{4cm}|}{\centering \textbf{Improvement rate}} & \multicolumn{1}{p{4cm}|}{\centering \textbf{Average improvement} [kcal/mol]} & \multicolumn{1}{p{4cm}|}{\centering \textbf{Improvement rate}} & \multicolumn{1}{p{4cm}|}{\centering \textbf{Average improvement} [kcal/mol]} &  \multicolumn{1}{p{4cm}|}{\centering \textbf{Default (adaptive)} [sec/molecule]} \\
        \hline
        aSTO-3G
        &  98.6\% & 8.56\,(9.53) & 82.9\% & 31.2 & 0.590 (0.590)\\
        a3-21G
        &  99.8\% & 39.9 & 96.7\% & 107.8 & 1.167 (1.167)\\
        a6-31G
        &  99.2\% & 3.29 & 99.9\% & 11.9 & 1.204 (1.205)\\
        a6-31G*
        &  99.2\% & 2.37 & 93.7\% & 11.0 & 1.464 (1.465) \\
        \hline
    \end{tabular}
    }
    \caption{Hartree-Fock total energy decrease (Left), atomization energy improvement vs 6-311G(2df,2pd) (Middle) and average computational cost comparison when employing adaptive vs default basis sets for 30,000 QM9 molecules (Right). 
    For aSTO-3G, the improvement in the total energy was additionally computed using optimal (non-ML) scaling factors and are given in brackets. 
    All timings evaluated on a compute node equipped with a 16-core AMD Ryzen 9 7950X @ 5.7GHz processor and 128 GB DDR5 RAM.}
    \label{tab:energies1}
\end{table*}
These metrics are quantified in Table \ref{tab:energies1} which shows changes in the energetics obtained from using adaptive basis sets predicted by the ML model trained on the first 2000 smallest QM9 molecules.
For all 4 basis sets, the adaptive versions lead to lowering of the total HF energies for almost all of the 30,000 tested molecules.
Lowering of the total HF energy indicates a better approximation to the optimal single Slater-determinant ground state wavefunction due to the variational principle. 
The success rate is particularly high for double-zeta basis sets, which is likely due to the higher flexibility offered by having two scaling factors compared to STO-3G, where all valence orbitals are scaled equally. 
For STO-3G, we computed optimal scaling factors $\boldsymbol{\zeta}_{\text{opt}}$ for all 30k validation molecules and found that our ML model on average recovered ~90\% of the energy improvement (8.56/9.53 kcal/mol) that can be achieved using $\boldsymbol{\zeta}_{\text{opt}}$ obtained via optimization.
The corresponding changes in atomization energies are slightly less consistent likely due to basis set superposition errors (BSSE) which were not present in the training data since the $\boldsymbol{\zeta}_{\text{opt}}$ were obtained by minimizing molecular total energies.
Nevertheless, the magnitude of the improvements can be seen to be even larger than for total energies.
Throughout our work, we found particularly large improvements in energy for the 3-21G basis set.
A likely explanation for this is the small size of the basis set along with the previously mentioned flexibility offered by two separate scaling factors.
We note here that, since smaller basis sets yield the largest accuracy increase, adaptive basis sets should be beneficial to Hamiltonians used in quantum computers where only small basis sets are employed due to hardware constraints\cite{vqe_adaptive_basis}.
We also note again that the optimal scalings for polarization functions (Fig.~\ref{fig:lc}) were found to be harder to learn which can likely be improved with a more balanced training set selection or a ML model that accounts for the coupling between the scaling factors into account by learning them together.
Larger improvements from the HF calculations should be possible with more accurate polarization function scalings since these are known to have a drastic impact~\cite{jensen_review_2013}.
In Table \ref{tab:energies1}, we also include average CPU timings for the HF calculations using adaptive vs default basis sets.
For adaptive basis sets, this contains the scaling factor prediction time by the ML model and the HF calculation time.
As can be seen, the increased accuracy using adaptive basis sets can be obtained at virtually no additional cost.
\\
\begin{table*}[!htbp]
    \centering
    \scalebox{0.8}{
    \renewcommand{\arraystretch}{1.5}
    \begin{tabular}{|l|ccccccccc|}
        \hline
         & \multicolumn{9}{c|}{\textbf{Atomization energy error [kcal/mol]}} \\
         \hline
         \textbf{Molecule}\hspace{1.5mm} & \hspace{1mm}\textbf{STO-3G}\hspace{1mm}  & \hspace{1mm}\textbf{aSTO-3G}\hspace{1mm}  & \hspace{1mm}\textbf{3-21G}\hspace{1mm}  &  \hspace{1.5mm}\textbf{a3-21G}\hspace{1.5mm} & \hspace{1.5mm}\textbf{6-31G}\hspace{1.5mm} & \hspace{1mm}\textbf{a6-31G}\hspace{1.5mm} & \hspace{1mm}\textbf{6-31G*}\hspace{1.5mm} & \hspace{1mm}\textbf{a6-31G*}\hspace{1.5mm} & \hspace{1mm}\textbf{6-31G**}\hspace{1.5mm}\\
        \hline
        Aspirin & 221.1 & 218.6 & 229.5 & 190.4 & 182.2 & 172.7  & 23.4 & 17.2 & 11.5\\
        Glucose & 207.1 & 113.3 & 253.6 & 207.6 & 198.4 & 188.9  & 37.5 & 8.7 & 10.4\\
         MDMA & 7.6 & 68.6 & 224.5 & 159.3 & 177.2 & 168.2 & 26.0 & 15.0 & 9.4\\
        Mescaline & 37.5 & 109.7 & 248.9 & 181.6 & 197.9 & 185.6  & 31.3 & 16.8 & 11.2\\
        Metformin & 114.9 & 99.3 & 188.4 & 165.2 & 150.3 & 144.5  & 29.6 & 24.8 & 11.3\\
        Methylglyoxal conformer 1\, & 83.6 & 84.3 & 99.7 & 84.8 & 86.6 & 83.2  & 8.7 & 4.6 & 4.4\\
        Methylglyoxal conformer 2\, & 78.5 & 82.5 & 100.1 & 83.7 & 82.1 & 78.2  & 8.6 & 4.1 & 4.3\\
        Nicotine & 39.7 & 13.6 & 181.7 & 135.9 & 139.0 & 133.4  & 20.2 & 13.2 & 5.9\\
        Paracetamol & 123.2 & 118.0 & 181.8 & 155.4 & 136.5 & 132.8  & 23.4 & 17.7 & 9.1\\
        Resveratrol & 172.6 & 196.4 & 265.5 & 226.4 & 190.1 & 186.6  & 36.1 & 31.3 & 13.6\\
        Salicylic acid & 175.2 & 153.6 & 176.3 & 158.2 & 136.3 & 132.4  & 21.8 & 19.3 & 9.2\\
        Uric acid & 372.0 & 258.3 & 245.6 & 227.5 & 201.7 & 191.3 & 30.5 & 33.6 & 20.3\\
        \hline
        \textbf{Mean Error} & \textbf{136.1} & \textbf{126.4} & \textbf{199.6} & \textbf{164.7} & \textbf{156.1} & \textbf{149.4} & \textbf{24.8} & \textbf{17.2} & \textbf{10.0}\\
        \hline
    \end{tabular}
    }
    \caption{Counterpoise corrected Hartree-Fock atomization energy errors compared to 6-311G(2df,2pd) for a test set of 12 out-of-domain biologically/societally relevant molecules using adaptive and default Pople-style basis sets. 
    The Methylglyoxal conformers correspond to  the aldehyde group being on the same (1) and opposite side (2) of the keto group.}
    \label{tab:special_molecs_atomization_counterpoised}
\end{table*}
 \begin{figure}[!htbp]
          \centering           
          \includegraphics[width=\columnwidth]{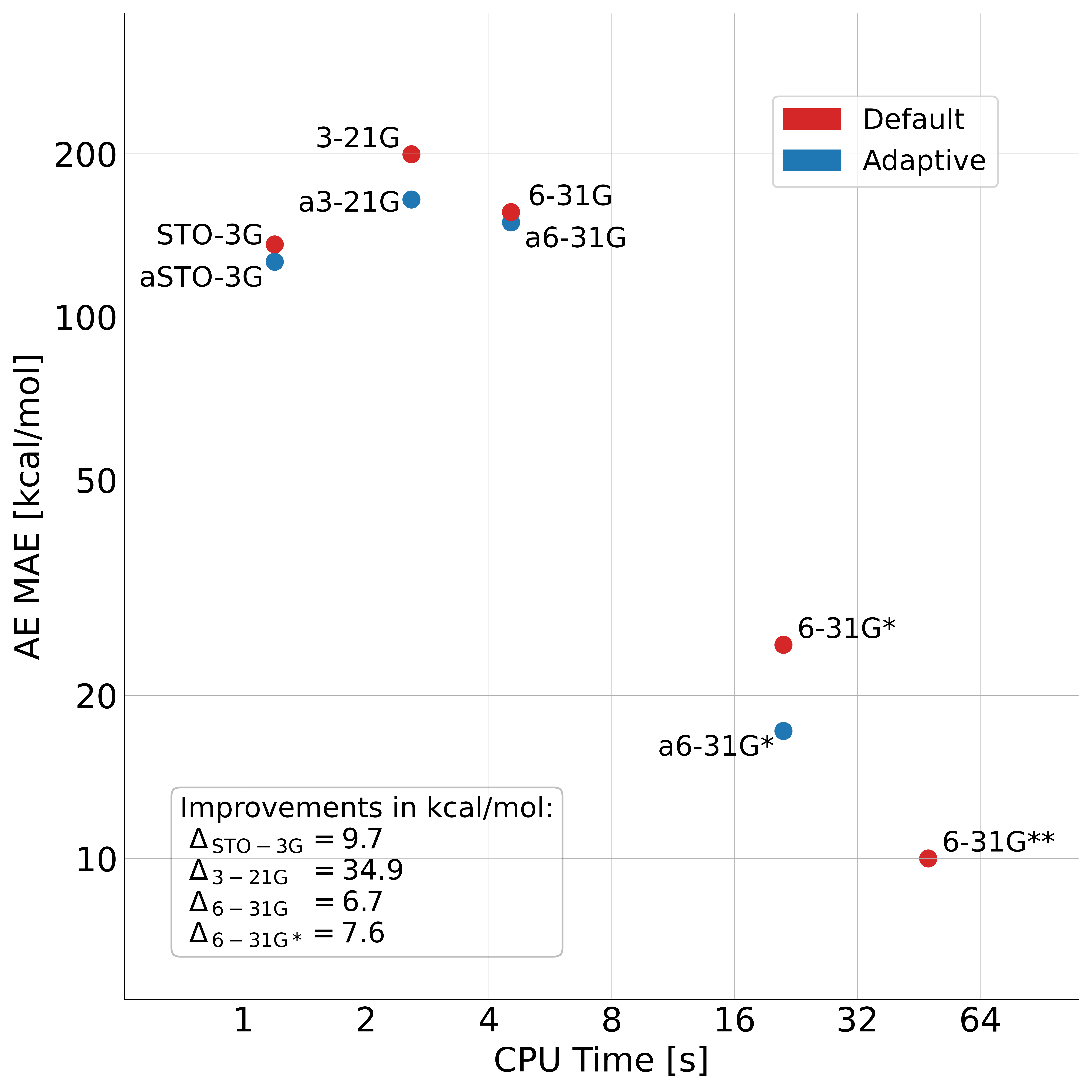}
          \caption{Hartree-Fock atomization energy mean absolute errors vs average computational cost for a test set of 12 biologically/societally relevant molecules (Table~\ref{tab:special_molecs_atomization_counterpoised}) using adaptive and default Pople-style basis sets. 
          All timings were evaluated on a compute node equipped with a 16-core AMD Ryzen 9 7950X @ 5.7GHz processor and 128 GB DDR5 RAM.}
     \label{fig:pareto}
 \end{figure}
To further test the transferability of the method we designed a test set of 12 larger, out-of-domain, biologically and societally relevant molecules.
All adaptive basis sets for these molecules were obtained through predictions from the ML model trained on the 2000 smallest QM9 molecules. 
Table 1 in the SI shows the change in total HF energy achieved using adaptive basis sets for these molecules while atomization energy errors are shown in table \ref{tab:special_molecs_atomization_counterpoised}.
Atomization energies of these molecules are compared directly to 6-311G(2df,2pd) and in order to mitigate BSSE, a counterpoise correction was applied to all basis sets. 
As expected, the error declines with increasing basis set size with the exception of STO-3G and for nearly all molecules, it consistently improves when adaptive basis sets are employed.
The better performance of STO-3G compared to 3-21G and 6-31G was verified using the Orca quantum chemistry package (version 5.0.3)\cite{orca5} and is likely linked to error cancellation between the total and atomic energies.
As mentioned earlier, the large impact of polarization functions can be seen as the a6-31G* basis set nearly reduces the error 10-fold compared to its upolarized form 6-31G.
The improvements obtained from each adaptive basis set in this case as well comes at virtually no added computational cost as shown in the computational cost vs accuracy plot in Fig.~\ref{fig:pareto}.
Similar to the observation from QM9 testing, particularly large improvement can be seen with the adaptive a3-21G basis but also with the inclusion of polarization functions in the a6-31G* basis set.
We expect to see a similar shift in the Pareto front for even larger and well balanced basis sets with more polarization functions being adaptable.
\\
\begin{table*}[!htbp]
    \centering
    \scalebox{0.75}{
    \renewcommand{\arraystretch}{1.1}
    \begin{tabular}{|l|cccc|cccc|}
        \hline
         & \multicolumn{4}{c|}{\textbf{HOMO-LUMO Gap [eV]}} & \multicolumn{4}{c|}{\textbf{Dipole Moment Norm [D]}} \\
         \hline
         \textbf{Molecule}\hspace{1.5mm} & \hspace{1mm}\textbf{3-21G/a3-21G}\hspace{1mm}  &  \hspace{1.5mm}\textbf{6-31G/a6-31G}\hspace{1.5mm} & \hspace{1mm}\textbf{6-31G*/a6-31G*}\hspace{1.5mm} & \hspace{1mm}\textbf{cc-pVQZ}\hspace{1.5mm} & \hspace{1mm}\textbf{3-21G/a3-21G}\hspace{1mm}  &  \hspace{1.5mm}\textbf{6-31G/a6-31G}\hspace{1.5mm} & \hspace{1mm}\textbf{6-31G*/a6-31G*}\hspace{1.5mm} & \hspace{1mm}\textbf{cc-pVQZ}\hspace{1.5mm} \\
        \hline
        Aspirin & 11.65/11.8 & 11.46/11.48  & 11.53/11.53 & 11.31 & 1.97/1.94 & 2.32/2.34 & 2.24/2.16 & 2.23 \\
        Glucose & 15.07/15.4 & 14.92/14.97  & 15.15/14.42 & 11.31 & 4.21/4.17 & 4.49/4.43  & 3.85/3.83 & 3.46 \\
        MDMA & 11.99/12.3 & 11.93/11.85  & 11.81/11.8 & 11.03 & 1.56/1.54 & 1.79/1.66  & 1.22/1.24 & 1.06 \\
        Uric Acid & 11.54/11.6 & 11.39/11.42 & 11.58/11.6 & 10.89 & 3.36/3.43 & 3.50/3.56  & 3.42/3.52 & 3.55 \\
        Mescaline & 12.17/12.5 & 12.11/12.08  & 12.1/12.1 & 10.8 & 5.53/5.35 & 6.03/5.47  & 4.71/4.52 & 4.24 \\
        Metformin & 13.32/13.55 & 13.06/13.1  & 13.44/13.45 & 11.56 & 4.47/4.32 & 4.56/4.52  & 4.45/4.45 & 4.38 \\
        Nicotine & 12.68/12.99 & 12.46/12.43  & 12.32/12.3 & 11.86 & 3.32/2.96 & 3.63/3.42  & 3.08/3.04 & 3.04 \\
        Paracetamol & 11.59/11.92 & 11.56/11.45  & 11.49/11.43 & 10.54 & 2.63/2.62 & 2.88/2.89  & 2.57/2.57 & 2.63 \\
        Resveratrol & 9.47/9.5 & 9.4/9.36 & 9.41/9.4 & 9.18 & 3.96/3.82 & 4.11/4.08  & 3.69/3.67 & 3.52 \\
        Salicylic Acid & 11.43/11.62 & 11.28/11.26  & 11.29/11.25 & 11.04 & 3.32/3.27 & 3.83/3.81  & 3.38/3.31 & 3.35 \\
        \hline
        \textbf{MAE} & \textbf{0.87/1.02} & \textbf{0.74/0.71} & \textbf{0.79/0.71} & \textbf{-} & \textbf{0.38/0.32} & \textbf{0.58/0.47}& \textbf{0.15/0.12} & - \\
        \hline
    \end{tabular}
    }
    \caption{HOMO-LUMO gaps and dipole moment norms using default and adaptive Pople-style basis sets.
    Mean absolute errors (MAE) are compared to the cc-pVQZ values in all cases.}
    \label{tab:special_molecs_other_props}
\end{table*}

Since an improvement in the wavefunction should correspond to an improvement in accuracy for all quantum mechanical properties, we further tested the changes in HOMO-LUMO gaps and dipole moments.
This is displayed in Tab. \ref{tab:special_molecs_other_props} for the same set of out-of-domain molecules.
All values are compared to the larger and more balanced cc-pVQZ basis set\cite{dunning1989gaussian} since dipole moment values are known to be highly sensitive to basis set size\cite{dipmom_polar_benchmark}.
All adaptive basis sets here as well were predicted by the ML model trained on the first 2000 smallest QM9 molecules.
As extensively discussed in prior studies, the form and choice of the optimal basis set can vary significantly for different molecular properties\cite{atmzn_energy_basis_set, property_basis_set1}. 
For both HOMO-LUMO gaps and dipole moments the errors do not consistently decrease with increasing basis set size, however the adaptive version improves the errors in all cases except the HOMO-LUMO gaps for a3-21G.
The dipole moment norms with a3-21G however do show an improvement indicating more accurate electron densities which is seen with all 3 adaptive versions.
Improvement of multiple properties that were not part of training highlights the "foundational model" aspect\cite{apbe0} of this method for basis sets.
This can also be exploited for the generation of accurate quantum chemistry datasets for ML while retaining the cost of a cheaper basis set similar to how the revQM9 dataset was generated using an adaptive exchange-correlation density functional\cite{apbe0}.  

\subsection{Transferability}
 \begin{figure*}[!htbp]
          \centering           
          \includegraphics[width=\linewidth]{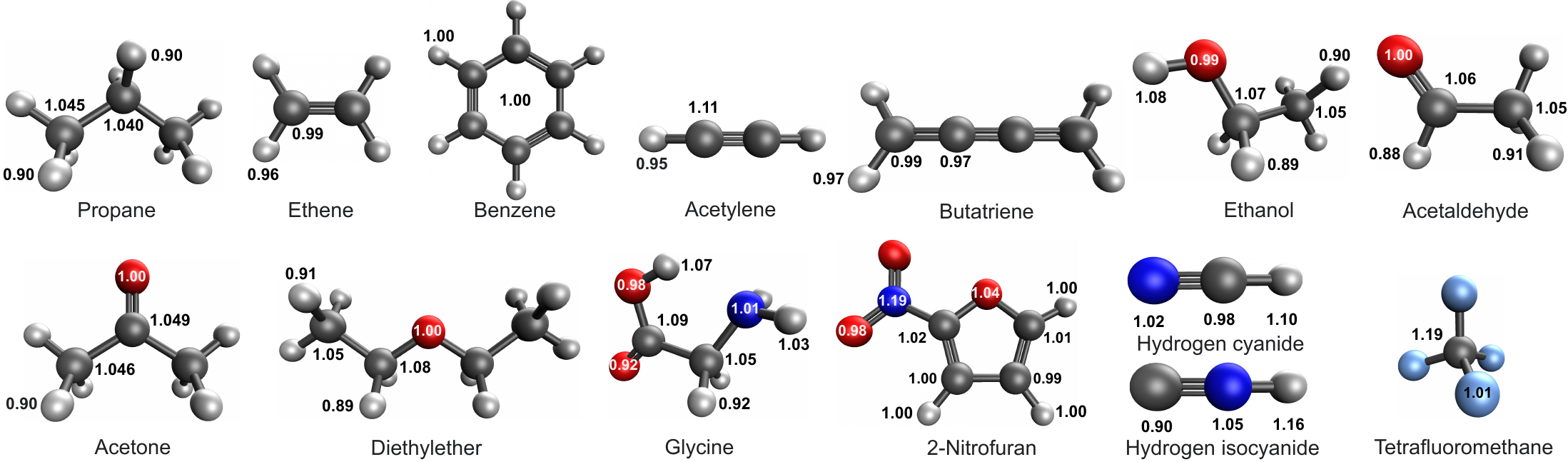}
          \caption{Exemplary molecules from the QM9 dataset with diverse functional groups and their optimal (HF energy-minimized) STO-3G valence orbital scaling factors $\zeta_\text{opt}$.}
     \label{fig:example_factors}
 \end{figure*}
The ease with which the orbital scaling factors can be learned, as indicated by our numerical results, suggests a strong mapping between them and an atom's local environment.
This is desirable since the ML model can be trained on a small dataset of diverse local chemical environments and still remain transferable and applicable to larger compounds.
Fig. \ref{fig:example_factors}, shows scaling factors for 14 molecules from the QM9 dataset with diverse functional groups and different orbital hybridizations for the STO-3G basis set.
Upon examination, $\zeta$ appears to be a very local quantity, for the most part influenced only by its direct bonding partners.
This is consistent with the local and short-ranged nature of atomic orbitals and underlines the flexibility of this approach. 
This strong locality allows partitioning large molecules with multiple functional groups into smaller fragments (or amons~\cite{amons_slatm}) with, for the most part, independent scaling factors.
This is likely the reason why our ML model seems to be transferable to larger molecules.
\\
In previous work, $\Delta$-\cite{delta2015} and multilevel\cite{bing_multi, m3l} machine learning methods have been used to predict corrections to calculations performed in small basis sets with significant improvements.
These improvements are even larger since molecular properties are learned from reference data using a larger basis set however separate models and labels are required for each property.
Adaptive basis sets, on the other hand, are a foundational ML model thereby providing improved wavefunctions and multiple improved properties/observables at once since the method remains fully \textit{ab initio}.
The training data acquisition cost is also lower since labels are generated via application of the variational principle to the same small basis set without need for more expensive, larger basis set reference calculations. 
Furthermore, the method might still be more transferable than $\Delta$-/multilevel methods due to the strong locality of the scaling factors depending on the functional groups.
Such comparisons will be analyzed in a future work employing a larger and more diverse training dataset while extending the method to basis sets from other families.


\section{Conclusion}

In this work have we have presented a  ML model for adaptive basis sets which improves Hartree-Fock wavefunctions at negligible additional cost.
The method predicts optimally scaled radial exponents of contracted Gaussian type orbitals for each atom within a molecule depending on its local chemical environment.
Optimal scaling factors for the STO-3G, 3-21G, 6-31G and 6-31G* Pople basis sets were generated via variational minimization of Hartree-Fock total energies.
Application of the variational principle provides training data without need for higher level reference calculations and guarantees simultaneous improvement of all observables due to more accurate ground state wavefunctions.
Testing on 30,000 random molecules from the QM9 dataset\cite{qm9} showed that the default basis sets could be improved upon already after training on only $\sim$100 molecules by using a local, atomic regression model (cf. Fig.\ref{fig:perf_curve}). 
After increasing the training set size to 2000 molecules, HF total and atomization energies systematically improve and the accuracy increase was found to be particularly large for 3-21G (cf. Tab \ref{tab:energies1}).

Making use of the high locality of $\zeta$, we investigated a variety of biologically/societally relevant molecules, most of which are significantly larger than the molecules used during training.
The model is found to be transferable, as shown by the observed improvements in energetics, HOMO-LUMO gaps and dipole moment norms for these larger molecules further highlighting a foundational model aspect (cf. Tabs. \ref{tab:special_molecs_atomization_counterpoised},\ref{tab:special_molecs_other_props}).
Analysis of the scaling factors for STO-3G indicates their strongly local character as similar functional groups and atoms with the same bonding partners often evoke similar optimal scaling factors (cf. Fig. \ref{fig:example_factors}).
This supports the locality hypothesis which underpins the  transferability of the method which is desirable in order to efficiently cover diverse chemistries through the use of suitable partitioning schemes, e.g.  amons\cite{amons_slatm} or GEMS~\cite{unke2024biomolecular}. 
Alternative methods for parameterizing $\zeta$ could be possible, 
however due to the increasing complexity of the problem for larger basis sets, 
we have opted for simple kernel ridge regression based machine learning with a compact local representation, such as many-body distribution functionals (MBDF).

While this work has focused on HF in order to exploit its variational nature,  scaled basis sets are expected to offer improvements to post-HF methods (such as CCSD(T) which is known to converge slowly with basis set size) building upon the improved ground state HF wavefunction, and also Kohn-Sham density functional theory based methods employing the same self-consistent-field framework as HF.
Since the size of the basis set remains constant within the adaptive basis set framework, improvements in wavefunction and properties of test molecules are obtained for negligible computational overhead, 
improving the computational cost vs accuracy trade-off for practically any post-Hartree-Fock quantum chemistry calculation (cf. Tab. \ref{fig:pareto}).
Regarding density functionals, note that  adaptive atomic basis sets can easily be paired with other adaptive parameters in the exchange correlation potential, such as the adaptive hybrid density functional approximation\cite{apbe0}.

Finally, while this work has been limited to the smallest Pople basis set family, the approach is expected to be equally applicable to   other popular basis set families in quantum chemistry, since they are also based on  similar contracted Gaussian type orbitals.
It is also straightforward to incorporate other chemical elements and to atomic orbital based treatment of condensed phase systems through programs, such as FHI-AIMS\cite{fhi-aims}.

\section{Data and code availability}
Generated training data, scripts and a trained model for obtaining optimal scaling factors for the basis sets discuessd in this work are available at \hyperlink{https://github.com/dkhan42/adaptive_basis}{https://github.com/dkhan42/aBasis}.
Script for generating MBDF along with gradients is available at \hyperlink{https://github.com/dkhan42/MBDF}{https://github.com/dkhan42/MBDF}.

\section{Acknowledgments}
D.K. acknowledges discussions with G. Domenichini. 
We acknowledge the support of the Natural Sciences and Engineering Research Council of Canada (NSERC), [funding reference number RGPIN-2023-04853]. Cette recherche a été financée par le Conseil de recherches en sciences naturelles et en génie du Canada (CRSNG), [numéro de référence RGPIN-2023-04853].
This research was undertaken thanks in part to funding provided to the University of Toronto's Acceleration Consortium from the Canada First Research Excellence Fund,
grant number: CFREF-2022-00042.
O.A.v.L. has received support as the Ed Clark Chair of Advanced Materials and as a Canada CIFAR AI Chair.
O.A.v.L. has received funding from the European Research Council (ERC) under the European Union’s Horizon 2020 research and innovation programme (grant agreement No. 772834).

\bibliographystyle{apsrev4-1}
\bibliography{literature.bib}

\end{document}